\documentclass[preprint,12pt]{elsarticle}
\usepackage{textcomp, gensymb}
\usepackage{amssymb,amsmath}
\usepackage{color,xcolor,todonotes}
\usepackage{graphicx,floatpag,fancyhdr}

\counterwithout{figure}{section}

\journal{}

\begin{document}

\begin{frontmatter}

\title{Analyzing stripes in crossing pedestrian flows using temporal matrices and a geometric model}

\author[label1]{Piotr Nyczka}
\ead{piotr.nyczka@pwr.edu.pl}
\author[label2]{Pratik\corref{cor2} Mullick}
\cortext[cor2]{corresponding author}
\ead{pratik.mullick@pwr.edu.pl}

\affiliation[label1]{organization={Department of Mathematics, Wrocław University of Science and Technology},
            addressline={Wyb. Wyspiańskiego 27}, 
            city={50-370},
            postcode={Wrocław}, 
            state={Lower Silesia},
            country={Poland}}

\affiliation[label2]{organization={Department of Operations Research and Business Intelligence, Wrocław University of Science and Technology},
            addressline={Wyb. Wyspiańskiego 27}, 
            city={50-370},
            postcode={Wrocław}, 
            state={Lower Silesia},
            country={Poland}}

\begin{abstract}

Understanding pattern formation in crossing pedestrian flows is essential for analyzing and managing high-density crowd dynamics in urban environments. This study presents two complementary methodological approaches to detect and characterize stripe formations, an emergent structure observed when two pedestrian groups cross at various angles. First, we propose a matrix-based method that utilizes time-resolved trajectory data to determine the relative crossing order of pedestrians from opposing groups. By identifying points of minimal spatial separation between individuals and analyzing associated time differences, we construct a crossing matrix that captures the sequence and composition of stripes. Second, we introduce a geometric model based on elliptical approximations of pedestrian groups, enabling analytical prediction of two key macroscopic quantities: the number of stripes and the interaction time between groups. The model captures how these quantities vary with the crossing angle and shows strong agreement with experimental data. Further analysis reveals that group elongation during crossing correlates with the vertical cross-section of the elliptical shape. These methods provide effective tools for analyzing large-scale movement datasets, informing the design of public spaces, and calibrating mechanistic models. The study also presents hypotheses about pattern transitions in continuous pedestrian streams, suggesting promising directions for future research on collective motion under varying flow geometries and densities.
\end{abstract}




\end{frontmatter}


\section{Introduction}
In modern urban environments, managing pedestrian and crowd dynamics \cite{wijermans2016landscape,feliciani2022introduction} has become a critical aspect of transportation research. As cities grow and urban spaces become more densely populated, the efficient movement of pedestrians and larger crowds plays an increasingly vital role in ensuring safety, accessibility, and overall functionality of public spaces \cite{bain2012living,cervero2017beyond,askarizad2020influence,wen2020higher}. Understanding crowd motion, particularly in crowded areas such as transit hubs \cite{hoy2016use,xu2022review}, stadiums \cite{haghani2018,gumbrecht2021crowds}, or public events \cite{zeitz2009crowd,gayathri2017review}, is crucial not only for optimizing flow but also for preventing dangerous overcrowding situations. Crowd motion studies have proven essential for developing better transportation systems \cite{gandhi2007pedestrian,haghani2018,haghani2020empirical}, as they provide insights into the collective behavior of individuals in dense environments \cite{moussaid2009collective,warren2018collective}, enabling planners to design infrastructure that accommodates large-scale pedestrian movement \cite{helbing2001,shahhoseini2019pedestrian}. These studies also inform the development of evacuation strategies in emergency scenarios \cite{guo2011collection,guo2012route,haghani2017stated}, ensuring rapid, safe movement under panic conditions \cite{chen2021extended,liang2021continuum}.

Crowd dynamics are particularly relevant in urban environments, where high pedestrian volumes interact with complex infrastructure, often leading to emergent patterns such as lanes \cite{johansson2008crowd,feliciani2016empirical,zhang2019pedestrian}, arches \cite{li2019arch,li2020fifty,frascaria2020emergent}, or stripe formations \cite{helbing2007,moussaid,helbing2012self,ploscb_pratik,worku2024detecting}. These dynamics are amplified in scenarios where multiple groups converge or cross paths, such as at intersections or transport terminals, where congestion can build quickly. For example, during peak hours, pedestrians navigating busy crosswalks may spontaneously form stripe-like patterns to avoid collisions with those moving in the opposite direction. Such emergent structures are not only fascinating from a scientific perspective but also have significant practical implications for traffic flow optimization \cite{moussaid,haghani2018,haghani2020empirical}, accident prevention \cite{THOMPSON1995131,johansson2008crowd,xie2021}, and the design of public spaces that minimize congestion and improve safety \cite{marcus1997people,southworth2005designing,helbing2005}. Additionally, understanding crowd behavior under panic, such as during evacuations, is critical for designing routes and strategies that minimize bottlenecks and improve the overall efficiency of emergency response systems.

The study of crowd dynamics in urban settings also benefits significantly from simulation-based research \cite{van2021algorithms}. Simulations offer a controlled way to model and analyze crowd behavior under various conditions 
, including overcrowded environments 
or during mass evacuations 
. By employing simulation techniques, one could potentially investigate how different factors, such as group size, crossing angles, or infrastructure design, affect pedestrian flows. These studies often rely on models such as agent-based simulations \cite{heliovaara2012counterflow,kountouriotis2014agent,alrashed2020agent} or cellular automata \cite{blue2001cellular,lu2017study}, which allow for detailed observation of local interactions between individuals. Despite the advances in simulation techniques, many models focus primarily on local behaviors, such as collision avoidance \cite{abdullah2009review,golas2013hybrid,charlton2019fast}, grouping within a crowd \cite{moussaid2010walking}, social pressure \cite{helbing2007}, but fail to capture the broader emergent structures that can arise from these interactions, such as the formation of stripes in crossing pedestrian flows \cite{helbing2005,ploscb_pratik,worku2024detecting}.

\begin{figure}[h!]
    \centering
    \includegraphics[width=\textwidth]{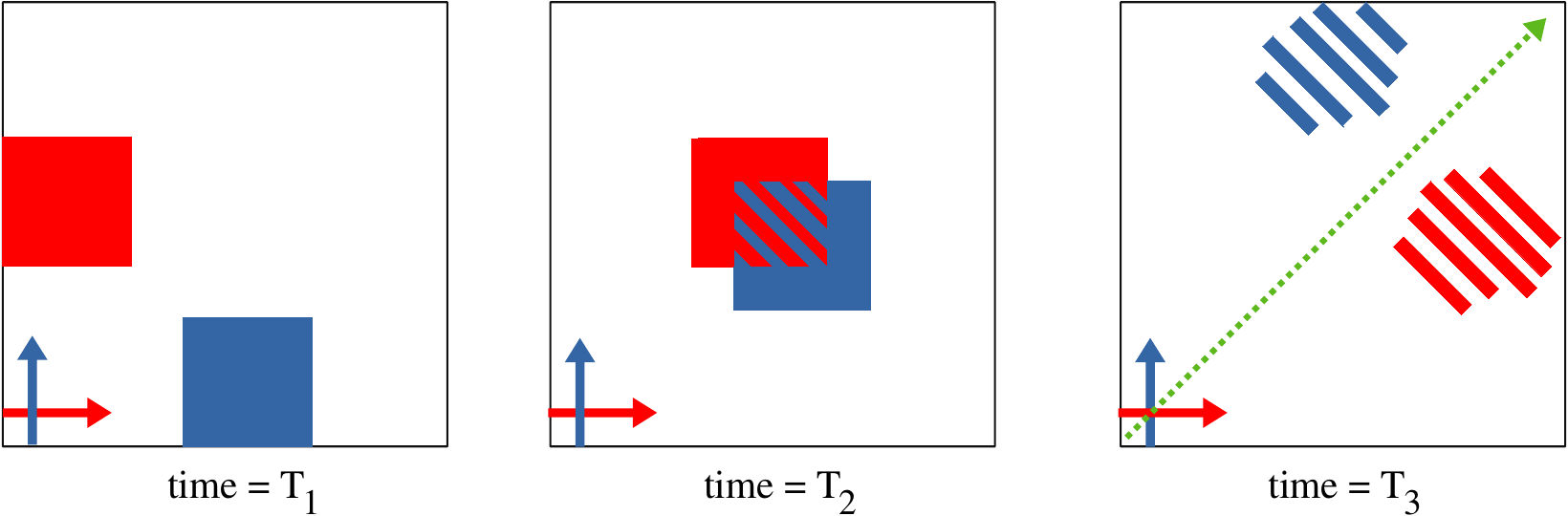}
    \caption{A schematic representation illustrating stripe formation as two groups of people cross paths. The figure presents three distinct moments: before crossing ($T_1$), during crossing ($T_2$), and after crossing ($T_3$), with $T_1<T_2<T_3$. The two groups are represented by blue and red squares, each moving in the direction indicated by matching colored arrows. A green dotted arrow marks the bisector of the crossing angle.}
    \label{fig:stripe_schematic}
\end{figure}

These emergent patterns, including lane and stripe formations, play a vital role in pedestrian dynamics, particularly in urban environments. Emergent behavior refers to complex patterns or structures that arise from simple individual actions without any central coordination. In the case of pedestrian movement, these patterns can lead to more organized flows, reducing collisions and improving overall efficiency. For example, stripe formations occur when two groups of pedestrians moving in opposite directions spontaneously align themselves into parallel lanes, allowing each group to move more smoothly with minimal interference. Formation of stripes for two crossing flows of pedestrians has been shown schematically in Figure \ref{fig:stripe_schematic}. This type of emergent behavior is especially important in scenarios where pedestrian groups cross paths, such as in busy intersections or plazas. Studying these emergent phenomena provides valuable insights into how large crowds behave in real-world environments and helps improve transportation systems' design and management \cite{ZANLUNGO1,ZANLUNGO2}.

Despite the importance of understanding pedestrian flow and emergent patterns, accurately modeling and analyzing such complex, dynamic systems presents significant challenges. Traditional models of pedestrian movement often rely on trajectory data to capture the paths of individuals over time \cite{bisagno2017data,amirian2019data,van2021algorithms}, which, while useful for certain analyses, may fail to detect larger emergent structures such as stripes. This is particularly true when multiple groups of pedestrians cross paths at various angles \cite{helbing2001,helbing2005,pedinteract_cecile,ploscb_pratik,worku2024detecting}, creating interactions that are not easily captured by trajectory-based methods alone. The difficulty of analyzing these situations highlights the need for more advanced models that can account for the global behaviors that emerge from local interactions, especially in crowded urban environments where crossing flows are common.


Various types of pedestrian flows exhibit different dynamic characteristics, depending on factors like crowd density, crossing angle, and the structure of the surrounding environment. For example unidirectional flows \cite{seyfried2005,zhang2013empirical,hu2021social}, where all pedestrians are moving in the same direction, are relatively easy to manage, as individuals naturally form lanes that increase flow efficiency. However, bidirectional flows \cite{guo2016uni,qin2018collective}, where groups move in opposite directions, or more complex scenarios where groups are crossing at angles \cite{helbing2001,helbing2005,pedinteract_cecile,ploscb_pratik,worku2024detecting}, are more challenging to model and analyze due to the greater likelihood of collisions and congestion. It is in these crossing flows, particularly in urban settings, where stripe formations are most commonly observed, making them a critical focus for transportation research.

Emergent structures like stripes are of particular interest because they represent a form of spontaneous order that enhances flow efficiency in situations where two groups of pedestrians interact. However, detecting and analyzing these formations using current methodologies remains a challenge, particularly in real-world environments where data may be incomplete or noisy, and computational efficiency is paramount. Traditional trajectory-based methods, while useful for tracking individual movements, often lack the sophistication needed to capture the full complexity of these patterns, especially in crowded or chaotic settings.

Another significant gap in the literature is the lack of computationally efficient techniques for analyzing large-scale pedestrian flows. While methods such as the edge-cutting algorithm can effectively detect stripe formations \cite{ploscb_pratik}, they are computationally intensive, making them less practical for real-time monitoring or large-scale urban applications. Furthermore, algorithms such as the pattern-matching technique \cite{ploscb_pratik,worku2024detecting} may not always provide consistent results, particularly when faced with variations in the quality of the input data. 

Given these limitations, the development of new methodologies that can efficiently detect and analyze stripe patterns in crossing pedestrian flows is a critical area of research. Such methods should not only reduce computational costs but also provide deeper insights into both individual and group dynamics, allowing for a more comprehensive understanding of how these patterns form and evolve. By addressing these gaps, researchers can advance the field of pedestrian flow modeling and contribute to the design of more efficient, safe, and optimized transportation systems. 

\subsection{Our contribution}
The main contribution of this paper is the development of two distinct yet complementary methods for detecting and analyzing stripe formation in crossing pedestrian flows: a matrix-based method and a barycentric (BC) approach.

Our matrix-based method, computationally efficient and fundamental in its design, considers microscopic-level interactions between individual pedestrians from opposing groups. Specifically, we compute the time difference in crossing for pairs of pedestrians from the human-trajectory data and use this information to construct a temporal matrix that encapsulates key aspects of stripe formation. This matrix not only reveals the number and size of stripes but also provides insights into the identity of individuals within each stripe and the sequence in which these stripes cross paths. 

On the other hand, the barycentric approach is built on transforming pedestrian trajectories into moving reference frames defined by the barycenters (centers of mass) of the entire system or individual groups. This perspective reveals the spatio-temporal structure of the crowd during crossing, highlighting stripe orientation, group deformation, and compression in an interpretable way. As part of this approach, we develop a geometric model that approximates pedestrian groups as ellipses and provides analytical predictions for the number of stripes and the interaction time as functions of the crossing angle. We find that group elongation during crossing is strongly correlated with the vertical cross-section of the ellipse, which in turn governs stripe formation. To extract stripe counts directly from barycentric views, we introduce two estimation strategies: the precise BCP method, which adapts thresholding and time selection per angle, and the simplified BCS method, which uses fixed parameters to offer a consistent and efficient alternative.

Finally, based on our findings, we propose two testable hypotheses regarding the emergence of traveling stripe waves in continuous pedestrian streams and the existence of a transition to disordered regimes when group widths exceed a critical threshold. These ideas lay the foundation for future theoretical and experimental research.


The remainder of this paper is structured as follows: in Section \ref{sec:experiments}, we describe the experimentally obtained trajectory data used for this analysis. Section \ref{sec:matrix_method} introduces the crossing matrix method and and section \ref{sec:barycenter} presents the barycentric reference frame formulation. In section \ref{sec:geometric} we develop a minimalist geometric model of elliptical pedestrian groups to predict number of stripes and crossing time as functions of the crossing angle. Section \ref{sec:number_of stripes} introduces two algorithms under the barycentric analysis framework to estimate the number of stripes based on transformed group trajectories. Section \ref{sec:res_dis} presents and discusses the results and finally section \ref{sec:conclusion} concludes the paper with a summary of key findings, broader implications, and suggestions for future research.
	
\section{Methodology}\label{sec:method}

\subsection{Experimental details}\label{sec:experiments}

This research utilized experimental data \cite{pedinteract_cecile,ploscb_pratik} on the crossing flows of pedestrians, obtained from live participants on the campus of University of Rennes, France. The experiments were conducted over two days, with 36 and 38 participants on the first and second days, respectively. The participants were divided into two groups with similar spatial densities and instructed to move along a direction announced before each trial, such that the two groups would cross each other at a particular angle. The experiments were performed for seven different crossing angles: $0\degree$, $30\degree$, $60\degree$, $90\degree$, $120\degree$, $150\degree$, and $180\degree$, with approximately 17 trials performed at each angle, totaling 116 trials. During each trial, the head trajectory of each pedestrian was recorded as a time series for approximately 15-25 seconds using VICON, an infrared camera system. The positions of the pedestrians were recorded at 120 Hz, and the center of the tracking area was used as the origin of a two-dimensional Cartesian coordinate system. The participants were not visually or locomotively impaired and were recruited using a mailing list for volunteers on the campus. The experiments were performed in a rectangular hall with taped visual references for facilitating the direction of motion. The data obtained from the experiments were low-pass filtered using a forward-backward 4th-order butterworth filter to reduce gait-induced oscillations. The analysis presented in this paper used the filtered data.

\subsection{Constructing the crossing matrix}\label{sec:matrix_method}

In this section we summarize the mathematical formulation of constructing the matrix, that has the potential to provide complete information about stripe formation. 

\begin{figure}[h!]
    \centering
    \includegraphics[width=0.5\textwidth]{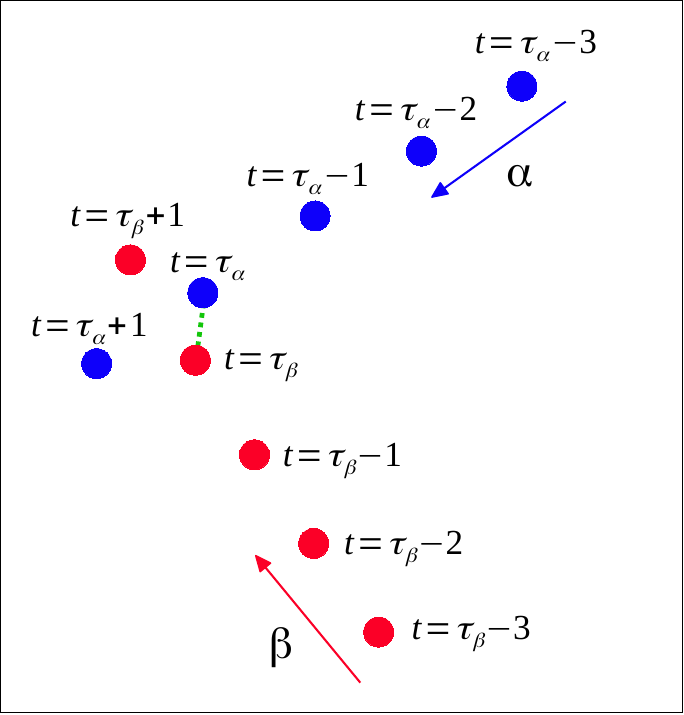}
    \caption{Schematic diagram showing the points of minimum distance between the trajectories of two pedestrians The two pedestrians are denoted by $\alpha$ and $\beta$, whose trajectories are denoted by blue and red circles respectively. The minimum distance is denoted by the green dotted line. The time-stamps of $\alpha$ and $\beta$ at these points are $\tau_\alpha$ and $\tau_\beta$ respectively.}
    \label{min_dist}
\end{figure}

For clarity in our discussion, we designate the two groups of crossing pedestrians as group $A$ and group $B$. We then consider a pair of pedestrians, $\alpha$ from group $A$ and $\beta$ from group $B$. Ideally, we are interested in the intersection point of the trajectories of these two pedestrians and then look at the respective time-stamps of the individuals crossing this point. However, the time-series human-trajectory data that we use for this research was recorded at time intervals of $1/120$ sec. For this reason, it is highly unlikely that the actual intersection point would have a time-stamp at all. So instead, we consider the points of minimum distance between the two trajectories. Note that this is not exactly the same as the \textit{distance of closest approach}, where both the individuals have the same time-stamp. The situation is described schematically in Figure \ref{min_dist}. 

In all the cases for our data, at the point of minimum distance between the trajectories of $\alpha$ and $\beta$, their respective time-stamps $\tau_\alpha$ and $\tau_\beta$ are different. Our goal is to find the time stamp $\tau_\alpha$ and $\tau_\beta$ such that
\begin{equation}
    (\tau_\alpha,\tau_\beta)=\underset{\tau_\alpha,\tau_\beta}{\operatorname{argmin}} \left\Vert P_\alpha (\tau_\alpha) - P_\beta (\tau_\beta) \right\Vert,
    \label{crossing_point}
\end{equation}
where $P_\alpha$ and $P_\beta$ are respectively the positions of $\alpha$ and $\beta$. If $\tau_\alpha<\tau_\beta$ we consider that `agent $\alpha$ crosses first with respect to agent $\beta$' and vice-versa. Subsequently, we create a matrix $M$ using the time-difference $\Delta \tau$ between a pair of pedestrians passing their respective points of minimum distance.

\begin{figure}[h!]
\centering
\includegraphics[width=\textwidth]{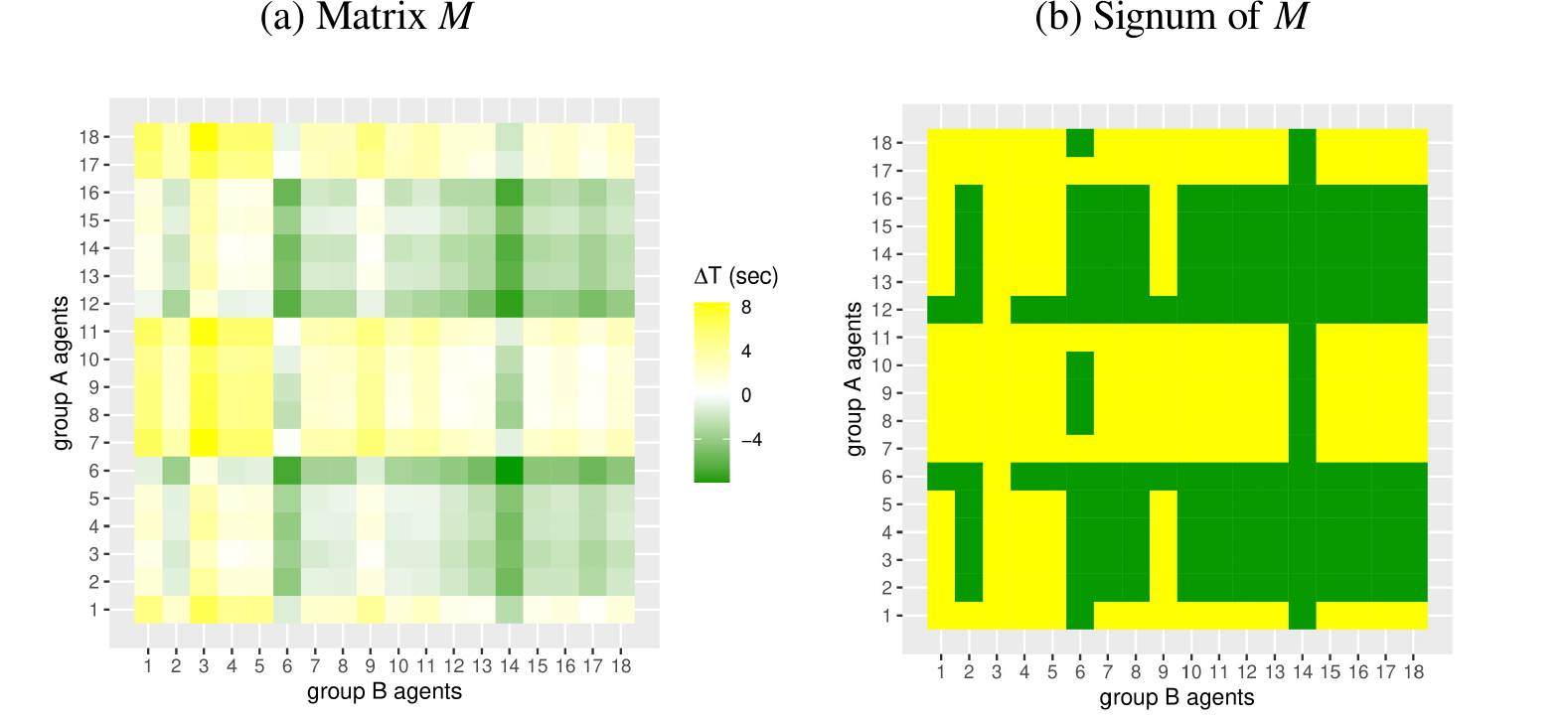}
\caption{The matrices $M$ and $\text{sgn}(M)$ for a typical configuration of crossing flows, where each group had 18 pedestrians. (a)The elements of the matrix $M$ are the time-differences $\Delta T$ for a pair of pedestrians, the values of which are represented by the color gradient on the right. (b) The matrix $\text{sgn}(M)$ has elements either $1$ or $-1$, denoted by yellow and green colors respectively. }
\label{matrices}
\end{figure}

For a pair of pedestrians $\alpha$ and $\beta$, if $\tau_\alpha$ and $\tau_\beta$ denote the time-stamps while being located on their respective points of minimum distance (See Figure \ref{min_dist}), $\Delta \tau_{\alpha\beta}$ is defined as
\begin{equation}
  \Delta \tau_{\alpha\beta}=\tau_\alpha-\tau_\beta.
  \label{delta_t}
\end{equation}
If $\Delta \tau_{\alpha\beta}<0$, it means that the agent $\alpha$ crosses first with respect to the agent $\beta$, and vice-versa. Finally, the matrix elements $M_{\alpha\beta}$ are defined as 
\begin{equation}
    M_{\alpha\beta}=\Delta \tau_{\alpha\beta}.
\end{equation}
We then consider all possible pairs of pedestrians from the two groups and evaluate the elements of the matrix $M$. Naturally $M_{\alpha\beta}=-M_{\beta\alpha}$ and for this reason we are only interested in the matrix spanned by the pedestrians from one group along the row and the pedestrians from other group along the column. Precisely we use the convention that (i) in the notation $M_{\alpha\beta}$, agent $\alpha$ belongs to group $A$ \& agent $\beta$ belongs to group $B$ and (ii) let $N_A$ and $N_B$ be number of agents in group $A$ and group $B$ respectively. If we define row vector $\mathbf{V_r}_{\alpha}$ and column vector $\mathbf{V_c}_{\beta}$ as \begin{equation}
    \begin{aligned}
        \mathbf{V_r}_{\alpha}(M) &= [M_{\alpha 1}, M_{\alpha 2}, ..., M_{\alpha N_B}], \\
        \mathbf{V_c}_{\beta}(M) &= [M_{1\beta}, M_{2\beta}, ..., M_{N_A \beta}]^T,
    \end{aligned}
    \label{vector_definition}
\end{equation}
then $\mathbf{V_r}_\alpha$ corresponds to agent $\alpha$ of group A and $\mathbf{V_c}_{\beta}$ corresponds to agent $\beta$ of group B. 
The size of the matrix would be $N_A\times N_B$. In Figure \ref{matrices}(a) we show the matrix $M$ for a typical configuration of crossing flows. The shades of colors yellow and green denote the positive and negative values of $M_{\alpha\beta}$ respectively. In section \ref{sec:res_matrix} we  describe in detail how this matrix could be exploited to obtain stripe formation details.

Note that if we are only interested in identifying pedestrians crossing first with respect to pedestrians from the other group, it would be sufficient to consider the signum function of matrix $M$. Obviously, $\text{sgn}(M)$ has the same properties as $M$: If $\text{sgn}(M_{\alpha\beta})=-1$, agent $\alpha$ crosses first with respect to agent $\beta$ and vice-versa for $\text{sgn}(M_{\alpha\beta})=1$. In Figure \ref{matrices}(b) we show $\text{sgn}(M)$ for the same configuration that has been shown in Figure \ref{matrices}(a). Considering agent 11 from group $A$ we can see that it crosses first with respect to all the agents from group $B$, except agent 14. On the other hand, there is no agent in group $A$ that crosses first with respect to agent 14 from group $B$. In other words, agent 14 from group $B$ crosses first with respect to all the agents from group $A$.

We want to clarify that, when we refer to a ``crossing" between two agents, we are referring to the intersection of their respective trajectories. Even if in our experiments the agents do not physically cross paths in the same time and space, we still consider that a ``crossing" has occurred because we are studying the spatio-temporal relationships between their resulting trajectories. In section \ref{sec:res_matrix}, we demonstrate that the matrix $\text{sgn}(M)$ could be used to get the stripe formation information very quickly.

\subsection{Barycentric reference frames}\label{sec:barycenter}	
In this study, we introduce a simple yet novel data analysis approach that, while unconventional in the field of pedestrian dynamics, is deeply rooted in classical mechanics, and originally stems from the magnificent work of Polish Astronomer Mikołaj Kopernik \cite{Kopernik_Mikołaj_(1473-1543)_De}, written centuries ago. Instead of analyzing pedestrian trajectories in static reference frames, we adopt alternative reference frames based on barycentric (center-of-mass) transformations. This method allows us to investigate the dynamics of movement relative to the overall system and its subsystems, providing new insights into stripe formation in crossing pedestrian flows. Although previous studies have briefly considered moving reference frames \cite{bacik2023}, our work significantly expands on this concept by systematically applying barycentric transformations to pedestrian trajectories.


Our methodology is designed to expose movement characteristics that are often overlooked by traditional approaches. By shifting perspectives to barycentric reference frames, we aim to capture the relative motion of pedestrian groups and identify collective behaviors that contribute to stripe formation. Depending on the context, different barycentric frames can provide unique insights, either by analyzing the dynamics of the entire system using a global barycenter or by focusing on individual groups through their respective barycenters.

\begin{figure}
\centering
\includegraphics[width=\textwidth]{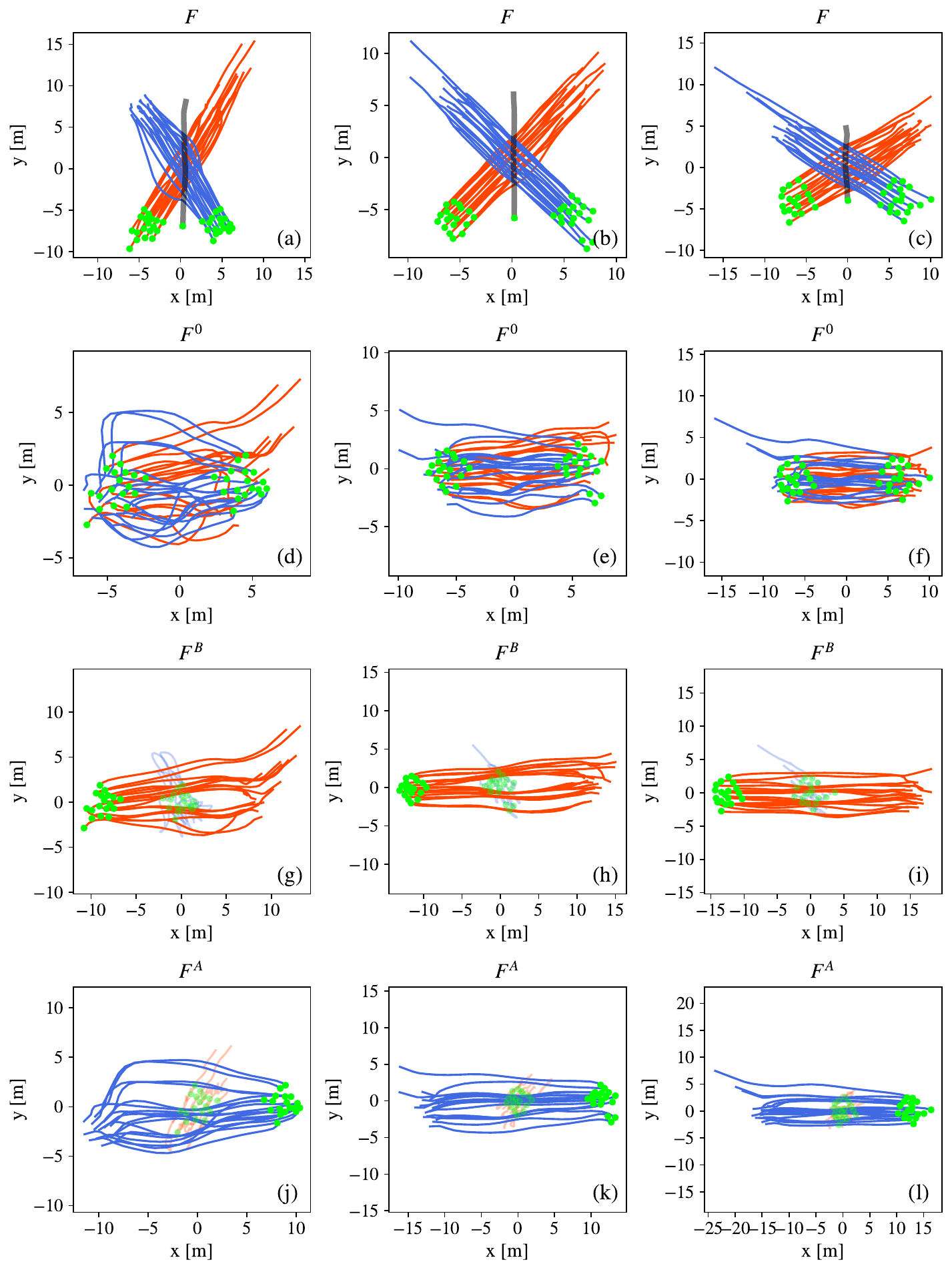}
\caption{Pedestrian trajectories in different reference frames, effectively decoding intricacies of the movement, shown for 3 typical trials belonging to different crossing angles. Columns denote respective crossing angles of $60 \degree$, $90 \degree$ and $120 \degree$. Panels (a), (b), (c) show trajectories in static reference frame, black line denotes trajectory of the global barycenter. Panels (d), (e), (f) show trajectories of groups $A$ and $B$ as seen from the group global barycenter. Panels (g), (h), (i) show trajectories of group $A$ as seen from group $B$'s barycenter. Panels (j), (k), (l) show trajectories of group $B$ as seen from the group $A$'s barycenter. The green dots indicate the starting points of the trajectories.}	
\label{fig:crossing-3-H}
\end{figure}

In our case, the system consists of two crossing groups of pedestrians, denoted as subsystems \( A = \{\alpha_i\} \) and \( B = \{\beta_j\} \), with group sizes \( N_A \) and \( N_B \), respectively. The positions of pedestrians in the group \( A \) are represented as \( \overrightarrow{\alpha_i} \), while those in group \( B \) are indicated as \( \overrightarrow{\beta_j} \). To analyze their dynamics, we introduce three key reference frames:  
\begin{itemize}
    \item \textit{Global barycentric frame} (denoted as \( F^0 \)) – The reference frame centered on the barycenter $C_0$ of the entire pedestrian system, capturing overall movement trends.
    \item \textit{Group-A-centric frame} (denoted as \( F^A \)) – A reference frame centered on the barycenter $C_A$ of group \( A \), allowing us to examine how group \( B \) moves relative to it.
    \item \textit{Group-B-centric frame} (denoted as \( F^B \)) – A reference frame centered on the barycenter $C_B$ of group \( B \), providing a perspective on how group \( A \) moves in relation to group \( B \).
\end{itemize} 
For clarity, we also define the static reference frame as $F$, in which the absolute positions of the barycenters are given by
\begin{eqnarray}
\overrightarrow{C_A}(t) &=& \frac{1}{N_A} \sum_{i=1}^{N_A} \overrightarrow{\alpha_i}(t) \nonumber \\ 
\overrightarrow{C_B}(t) &=& \frac{1}{N_B} \sum_{j=1}^{N_B} \overrightarrow{\beta_j}(t) \nonumber \\ 
\overrightarrow{C_0}(t)  &=& \frac{1}{2}(\overrightarrow{C_A}(t) + \overrightarrow{C_B}(t) ) 
\end{eqnarray}
In Figure \ref{fig:crossing-3-H} we have shown pedestrian trajectories in reference frames $F$, $F^A$ and $F^B$ for 3 typical trials from the crossing flows data set. Trajectories obtained in these reference frames are further analyzed, and in some cases additional post-processing steps are applied.

Consequently we can define positions of the pedestrians and the barycenters $C_u$ in $F^w$ as
\begin{eqnarray}
\overrightarrow{\alpha_i^w}(t) &=& \overrightarrow{\alpha_i}(t) - \overrightarrow{C_w}(t) \nonumber \\
\overrightarrow{\beta_j^w}(t) &=& \overrightarrow{\beta_j}(t) - \overrightarrow{C_w}(t) \nonumber \\
\overrightarrow{C_u^w}(t) &=& \overrightarrow{C_u}(t) - \overrightarrow{C_w}(t) 
\end{eqnarray}
where $u, w \in \{ 0, A, B \}$. Similarly, we define the velocity of barycenter $C_u$ in $F^w$ as
\begin{eqnarray}
\overrightarrow{V_u^w}(t) = \frac{\overrightarrow{d C_u^w}(t)}{dt}
\end{eqnarray}
and in the static reference frame $F$ as
\begin{eqnarray}
\overrightarrow{V_u}(t) = \frac{\overrightarrow{d C_u}(t)}{dt}
\end{eqnarray}


To ensure consistency across trials and simplify subsequent analysis, we rotated our coordinate system so that the average velocity of the global barycenter in the static reference frame $F$, denoted $\overrightarrow{V_0}(t)$, is aligned vertically upward. For technical accuracy, the velocity $\overrightarrow{V_0}(t)$ was averaged over the time interval from 25\% to 75\% of the total duration of a trial, thus minimizing transient effects at the beginning and end of the trial.

\subsubsection{Temporal phases of crossing dynamics}
\label{sec:phases}

When two groups of pedestrians cross each other, their interaction can be characterized by three key moments: the \textit{first contact}, the \textit{closest proximity}, and the \textit{last contact}. We denote these characteristic times as $\tau_1$, $\tau_2$, and $\tau_3$, respectively. These temporal markers capture the essential phases of the crossing process and are visually indicated by vertical lines in Figure \ref{fig:crossing-3-H}.


The \textit{first contact} $\tau_1$ is the moment when individuals from opposing groups begin to intermix spatially. Mathematically, this is defined as the earliest time $t = \tau_1$ at which at least one pedestrian $\alpha_i$ from the group $A$ is closer to the barycenter $C_B$ of the group $B$ than to its own group barycenter $C_A$, while simultaneously at least one pedestrian $\beta_j$ from group $B$ is closer to $C_A$ than to $C_B$. This condition could be summarized as follows,
\begin{eqnarray}
\exists_{i,j} \left[\left(\left\Vert\overrightarrow{\alpha_i^B}(t)\right\Vert < \left\Vert\overrightarrow{\alpha_i^A}(t)\right\Vert\right) \wedge \left(\left\Vert\overrightarrow{\beta_j^A}(t)\right\Vert < \left\Vert\overrightarrow{\beta_j^B}(t)\right\Vert\right)\right]
\label{eq:contact}
\end{eqnarray} Similarly, the \textit{last contact}  $\tau_3$ is the final moment when this condition holds. Thus, $\tau_1$ is defined as the smallest $t$ satisfying condition \eqref{eq:contact}, while $\tau_3$ is the largest $t$ for which this condition holds.


On the other hand, the \textit{closest proximity} $\tau_2$ occurs at the time when the distance between the barycenters of the two groups is minimized. This represents the moment when the two groups are most spatially compact around each other.
	
Thus, the three characteristic times $\tau_1$, $\tau_2$ and $\tau_3$ could formally be defined as
\begin{eqnarray}
\tau_1 &=& \underset{t}{\operatorname{argmin}} \left\{ \exists_{i,j} \left[\left(\left\Vert\overrightarrow{\alpha_i^B}(t)\right\Vert < \left\Vert\overrightarrow{\alpha_i^A}(t)\right\Vert\right) \wedge \left(\left\Vert\overrightarrow{\beta_j^A}(t)\right\Vert < \left\Vert\overrightarrow{\beta_j^B}(t)\right\Vert\right)\right]\right\} \nonumber \\
\tau_2 &=& \underset{t}{\operatorname{argmin}} \left\Vert \overrightarrow{C_A^B}(t)  \right\Vert \\
\tau_3 &=& \underset{t}{\operatorname{argmax}}\left\{ \exists_{i,j} \left[\left(\left\Vert\overrightarrow{\alpha_i^B}(t)\right\Vert < \left\Vert\overrightarrow{\alpha_i^A}(t)\right\Vert\right) \wedge \left(\left\Vert\overrightarrow{\beta_j^A}(t)\right\Vert < \left\Vert\overrightarrow{\beta_j^B}(t)\right\Vert\right)\right]\right\} \nonumber 
\label{eq:contact_times}
\end{eqnarray}

Finally, we define the \textit{total crossing time} $\tau^*$ as the duration over which the two groups remain intermixed, given by
\begin{eqnarray}
\tau^* = \tau_3 - \tau_1.
\end{eqnarray}

This definition of temporal phases provides a structured way to analyze the interactions of pedestrian groups, capturing both the initial mixing, the moment of greatest interaction, and the eventual separation. These time markers serve as critical indicators for understanding the nature of stripe formation and the evolution of crossing flows.

\subsection{A geometric model of crossing pedestrian flows}\label{sec:geometric}

To analytically understand aspects of the crossing phenomena and subsequent stripe formation, here we employ simple geometric arguments. As argued earlier, it is convenient to analyze the situation in the static reference frame $F$. Observations from experimental data indicate that each pedestrian group exhibits elongation in the direction of movement and that the boundaries of the group appear to be rounded. This was referred to as the `squeezing' in \cite{ploscb_pratik}. Our goal is to model the shape and dynamics of the groups in a way that accurately reflects these observations while maintaining simplicity.

\begin{figure}[h!]   
\centering
\includegraphics[width=\textwidth]{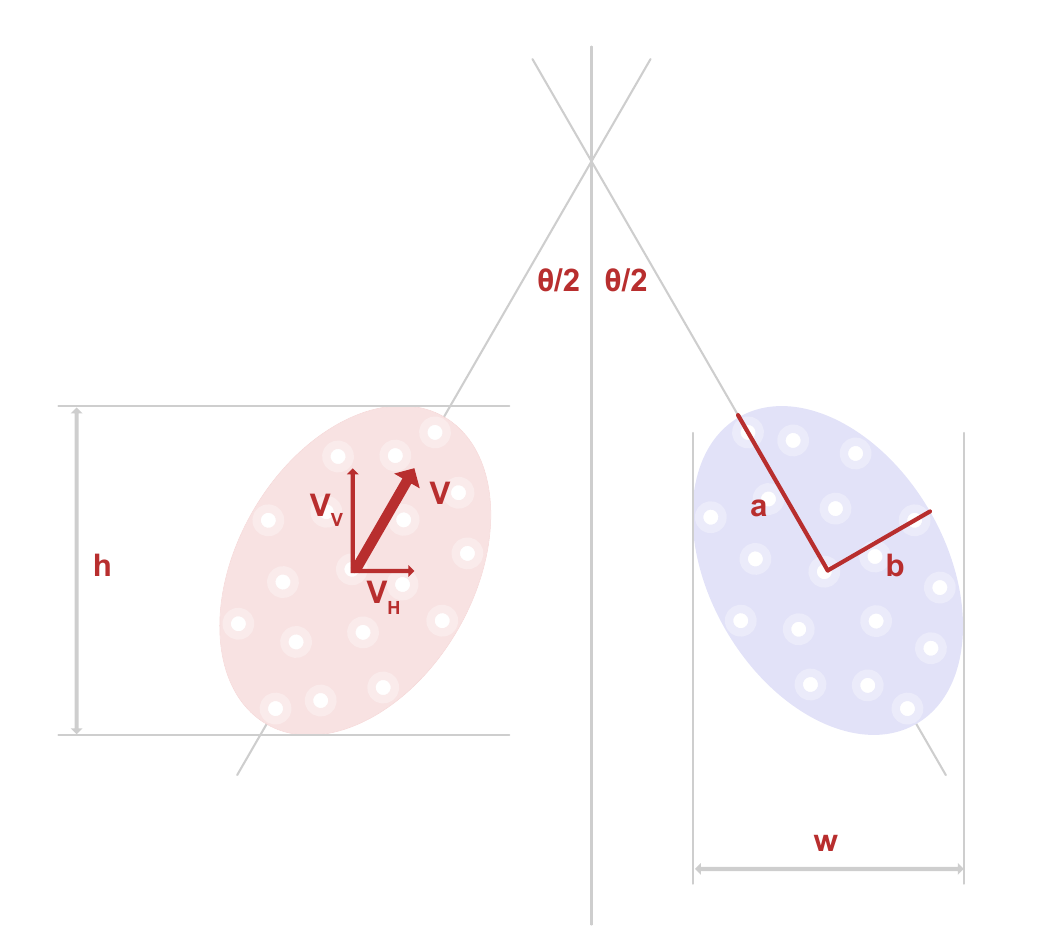}
\caption{Schematic representation of the theoretical model for crossing pedestrian groups. The two elliptical groups, are about to intersect at a crossing angle $\theta$. The velocity of each group is denoted by $V$, with its vertical and horizontal components represented as $V_V$ and $V_H$, respectively. The groups are modeled as ellipses with semi-major and semi-minor axes $a$ and $b$ respectively. The effective height $h$ and width $w$ of the groups are measured in a rotated reference frame, where the velocity of the global barycenter $C_0$ is aligned vertically.}	
\label{fig:ellipse}
\end{figure}

Among the most fundamental geometric shapes, squares and circles do not exhibit elongation, making them unsuitable for capturing the observed group structure. Rectangles, while elongated, lack rounded edges. In contrast, ellipses exhibit both elongation and smooth boundaries, making them a natural choice for modeling pedestrian groups. This is because elongation plays a crucial role in determining how groups interact as they cross each other. For simplicity, we assume that both groups have the same size and shape. Figure \ref{fig:ellipse} illustrates the geometric model within the reference frame $F$. 

In this setup, we define two groups, $A$ and $B$, crossing at an angle $\theta$. The velocity of each group is denoted by $V$, with vertical and horizontal components $V_V$ and $V_H$, respectively. The groups are modeled as ellipses, where $a$ and $b$ represent the semi-major and semi-minor axes. The effective height and width of the rotated ellipses are denoted as $h$ and $w$, respectively.

To facilitate analysis, we introduce a rotated reference frame in which the velocity of the barycenter of the system $\overrightarrow{V_0}(t)$ is oriented vertically. In this rotated frame, both groups move in opposite horizontal directions and cross each other over a well-defined time interval. This transformation simplifies the analysis by ensuring that the primary movement components align with the coordinate axes, making it easier to quantify the interactions between the groups. In the following, we discuss two important quantities related to the crossing behavior of the two groups and estimate them using our model.

\subsubsection{Number of stripes}

Using our minimalist model of crossing pedestrian groups, where each group is represented by an elliptical shape, we can estimate the number of stripes $\hat{n}$ that form in the crossing region. We assume that $\hat{n}$ is proportional to the effective height $h$ of the group in the global barycentric reference frame $F^0$, which represents the projected vertical dimension of the elliptical shape in the transformed reference frame. This assumption simplifies the analysis by completely ignoring the shape of the forefront of the crossing groups in $F^0$, focusing instead on how the vertical extent of the group governs stripe formation. As a result, we expect an angular dependency $\hat{n}(\theta)$, since the crossing angle $\theta$ directly influences the effective height $h$. This dependence was also previously discussed in \cite{ploscb_pratik}. Based on this assumption, the number of stripes could be expressed as:
\begin{eqnarray}
\hat{n} = S h = 2S \sqrt{a^2 \sin^2\frac{\theta}{2} + b^2 \cos^2\frac{\theta}{2} }
\label{eq:stripe_number}
\end{eqnarray} where $S$ is a proportionality constant related to the average minimum distance between pedestrians $\overline{d}_{min}$ within the group as $S = 1 / \overline{d}_{min}$. As $\overline{d}_{min}$ can be understood as an approximate width of a stripe, its reciprocal $S$ is the approximate number of stripes per length unit. The remaining expression on the right-hand side represents the effective height $h$ of an ellipse with semi-major and semi-minor axes $(a, b)$, rotated by an angle $\pi/2 - \theta/2$, as schematically illustrated in Figure \ref{fig:ellipse}.




\subsubsection{Crossing time}

The elliptical model of pedestrian crossing flows can also be used to estimate the crossing time, as defined in Section \ref{sec:phases}. To do this, we consider the horizontal velocity $V_H$ of the groups, which can be expressed as $V_H = w/t$, where $w$ is the effective width of the groups and $t$ is the corresponding time interval, as shown in Figure \ref{fig:ellipse}. 

Due to the symmetry of the crossing process, it is sufficient to analyze the movement of a single group. The crossing time is then defined as the duration between the \textit{first contact} and the \textit{last contact} of the barycenter of the group with the global barycenter $C_0$. Based on our elliptical model, the estimated crossing time $\hat{\tau}^*$ can be expressed as:
\begin{equation}
\hat{\tau}^* = \frac{2 \sqrt{a^2 \cos^2\frac{\theta}{2} + b^2 \sin^2\frac{\theta}{2} }}{V \sin\frac{\theta}{2}}
\label{eq:crossing-time}    
\end{equation} where the numerator represents the effective width of the rotated ellipse, capturing the horizontal extent of the group, and the denominator accounts for the horizontal projection of the velocity of the group, which governs the time required for the group to traverse the crossing region. This formulation provides a theoretical estimate of the crossing time, directly linking it to the geometric properties of the pedestrian groups and the crossing angle.

\subsection{Estimating the number of stripes from barycentric analysis} \label{sec:number_of stripes}

Here we introduce two variations of the barycentric method (BC) for determining the number of stripes: the precise method (BCP) and the simplified method (BCS). These approaches differ in how they handle algorithmic choices, with BCP focusing on optimizing performance for each crossing angle individually, while BCS prioritizes consistency and simplicity across all angles. Given that different algorithmic variants perform better for certain angles but worse for others, the BCP method tests multiple configurations and selects the most suitable one for each crossing angle separately. In contrast, BCS adopts a single fixed approach for all angles, sacrificing some precision for greater simplicity and robustness.

\begin{figure}[h!] 
\centering
\includegraphics[width=\textwidth]{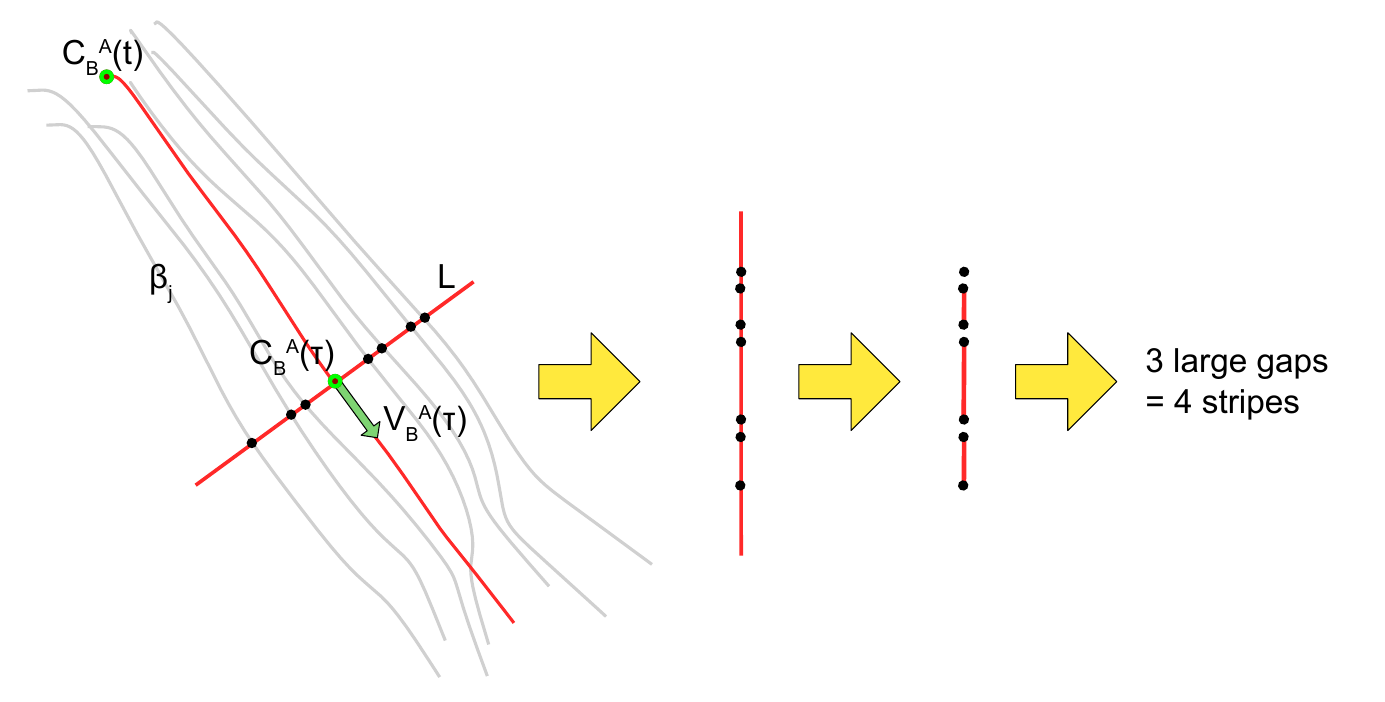}
\caption{Schematic illustration of extracting number of stripes from transformed trajectories. $\beta_j$ denote trajectories of pedestrians from group $B$. $C_B^A$ and $V_B^A$ are the trajectory and velocity of group B's barycenter as seen in $F^A$ (group A's barycentric frame). The line $L$ denotes a cross-section perpendicular to the velocity vector $\overrightarrow{V_A^B}$ and the black dots are intersections of $B_j$ with L. Further to the right we show the three prominent gaps identified in this cross section.}    
\label{fig:gaps-to-stripes}
\end{figure}

In both of these methods, the core idea is to analyze the trajectories of one pedestrian group from the perspective of the barycentric reference frame of the other group. As an example, we examine the trajectories of group \( B \), denoted as \( \beta_j \), as seen from the barycentric reference frame of group \( A \) (\( F^A \)). These transformed trajectories, denoted as \( \beta_j^A \), allow us to analyze the relative motion of the groups (see Figure \ref{fig:gaps-to-stripes}). To determine stripe formation, we analyze the cross section of these trajectories along a line \( L \), which is perpendicular to the velocity vector \( \overrightarrow{V_B^A}(\tau) \) of the barycenter of the group \( B \) as seen from \( F^A \) at time \( \tau \).
To systematically study the formation of stripes, we define a set $\{\tau_i\}$ of five characteristic time points where this cross-section is to be studied
\begin{equation}
\{\tau_i\} = \{ \tau_1, \tau_{1.5}, \tau_2, \tau_{2.5}, \tau_3 \}
\end{equation} where $\tau_1$, $\tau_2$ and $\tau_3$ have been defined earlier. \( \tau_{1.5} = \frac{1}{2}(\tau_1+\tau_2) \) and \( \tau_{2.5} = \frac{1}{2}(\tau_2+\tau_3) \) are intermediate time points that we additionally consider here. At each of these time points, the cross-section forms a one-dimensional distribution of points, as illustrated in Figure \ref{fig:gaps-to-stripes}. We then estimate the number of stripes by analyzing gaps between consecutive points in the cross-section.

To distinguish gaps between pedestrians within the same stripe from gaps separating different stripes, we assume that inter-stripe gaps are larger than intra-stripe gaps. We classify the gaps \( \{g_i\} \) using two different thresholding approaches:
\begin{itemize}
    \item \textit{absolute thresholding}: each gap is compared against a fixed threshold \( T_a \) 
    \item \textit{relative thresholding}: each gap is compared to the largest gap in the cross-section, scaled by a fractional threshold \( T_r \)
\end{itemize}
The corresponding expressions for estimating the number $n$ of stripes can be written as,
\begin{eqnarray}
n_a(\tau_i, T_a) &=& \sum{\{1 : g_i \geq T_a\}}_i + 1, \\
n_r(\tau_i, T_r) &=& \sum{\{1 : g_i \geq T_r \max{\{g_i\}}\}}_i + 1.
\end{eqnarray}
Here, \( n_a \) and \( n_r \) denote the number of stripes estimated using absolute and relative thresholding, respectively. The addition of 1 accounts for the fact that the number of stripes is always greater than the number of separating gaps by one. Since \( g_i \) depends on the time point \( \tau_i \), and \( n_a \) and \( n_r \) depend on \( T_a \) and \( T_r \) respectively, the estimated numbers of stripes are expressed as $n_a(\tau_i, T_a)$ and $n_r(\tau_i, T_r)$.

For each crossing angle, multiple time points may yield different values of \( n_a \) or \( n_r \). Instead of relying on a single time point, we consider all possible subsets of the 5 characteristic time points \( \{\tau_1, \tau_{1.5}, \tau_2, \tau_{2.5}, \tau_3\} \), denoted generically as \( \{\tau_i\} \). Given a set of time-dependent stripe estimates \( \{n_a(\tau_i, T_a)\} \) or \( \{n_r(\tau_i, T_r)\} \), we apply several statistical functions $f$ to obtain the final stripe count, viz.
\begin{itemize}
\setlength\itemsep{0.3cm}
 \item \textbf{max:} \( \max{\{n_z(t_i, T_z)\}} \)
 \item \textbf{min:} \( \min{\{n_z(t_i, T_z)\}} \)
 \item \textbf{ceil:} \( \left \lceil{\overline{\{n_z(t_i, T_z)\}}}\right \rceil \)
 \item \textbf{floor:} \( \left \lfloor{\overline{\{n_z(t_i, T_z)\}}}\right \rfloor \)
 \item \textbf{round:} \( \text{round}{\overline{\{n_z(t_i, T_z)\}}} \)
\end{itemize} where $z=\{a,r\}$. We exclude the arithmetic mean since the stripe count must be an integer.

For the precise method (BCP), we systematically optimize our algorithm for each crossing angle. In this case, we optimize the selection of the threshold type, time points, and processing function for each crossing angle separately. The algorithm could be described as follows:
\begin{enumerate}
    \item For each crossing angle, iterate over different values of \( T_z \)
    \item For each \( T_z \)
    \begin{enumerate}
        \item[i.] Select a subset of time points \( \{\tau_i\} \)
        \item[ii.] Compute \( \{n_z(t_i, T_z)\} \)
        \item[iii.] Apply a function \( f(\{n_z(t_i, T_z)\}) \) to estimate the number of stripes
        \item[iv.] Calculate the mean absolute error (MAE) between the predicted stripes and the reference edge-cutting (EC) algorithm \cite{ploscb_pratik}
    \end{enumerate}
    \item Select the optimal \( T_z \) that minimizes MAE
    \item Repeat the procedure for all possible subsets \( (2^5-6 = 26) \), functions ($5$), and threshold types ($2$), yielding $260$ possible combinations.
\end{enumerate} The variant with the lowest MAE is selected as the best for that crossing angle.

Now, we discuss the simplified approach, BCS. In this method, we select a single type of threshold and a single time point, making it more straightforward than the BCP. Since only one time point is chosen, there is no need to apply any statistical function $f$. For threshold selection, we opt for the absolute threshold $T_a$, as it is less sensitive to outliers in individual experiments. The choice of time point $\tau_i$ is based on the observed evolution of stripe formation. The stripes begin to emerge around $\tau_1$ and gradually fade beyond $\tau_3$. Although $\tau_2$ corresponds to the moment of greatest proximity between the groups, we observe that the maximum separation of stripes in the cross section occurs somewhere between $\tau_2$ and $\tau_3$. Based on this observation, we select $\tau_{2.5}$ as a balanced and effective choice.

In summary, while BCP ensures higher precision by adapting to each crossing angle, BCS offers a computationally simpler and more robust alternative. The choice of method depends on the application and whether precision or consistency is the priority. In Table \ref{tab:bcp_vs_bcs} we provide a comparative summary of these two methods.

\begin{table}[h!]
    \centering
    \caption{Comparative summary of BCP and BCS methods in estimating the number of stripes in the crossing region.}
    \begin{tabular}{|l|c|c|c|}
         \hline
         Method&Threshold&Time points&Optimization\\
         &type&used&approach\\
         \hline
         BCP (precise)&Absolute \& Relative&Multiple $\{\tau_i\}$&Selected per crossing angle\\
         &&&based on MAE\\
         BCS (simple)&Absolute&Single (\( \tau_{2.5} \))&Fixed across all angles\\ 
         \hline
    \end{tabular}
    \label{tab:bcp_vs_bcs}
\end{table}

\section{Results and Discussion}\label{sec:res_dis}

\subsection{Extracting insights from the matrices}\label{sec:res_matrix}

As described in section \ref{sec:matrix_method} each agent in the matrix $M$ or $\text{sgn}(M)$ is represented by a row vector $\mathbf{V_r}_\alpha$ or a column vector $\mathbf{V_c}_\beta$, and shows a unique pattern, as is evident from Figure \ref{matrices}. In this section we will demonstrate that agents with a similar pattern belong to the same stripe. Specifically, if we only consider $\text{sgn}(M)$, then agent $\alpha_1$ and agent $\alpha_2$ from group $A$ are of the same stripe if 
\begin{equation}
    \text{sgn}(\mathbf{V_r}_{\alpha_1}) = \text{sgn}(\mathbf{V_r}_{\alpha_2}),
    \label{eq_sgn_vector}
\end{equation}
and the same argument goes for agent $\beta_1$ and $\beta_2$ in group $B$. For instance, in Figure \ref{matrices}(b), agents 1, 8, 9, 10, and 18 of group $A$ are of the same strip, agents 1, 4, 5, and 9 are of group $B$ are of the same stripe, etc.

By definition, a stripe is known as a subset of pedestrians from one group that is not penetrated by pedestrians from the other group. In the edge-cutting algorithm \cite{ploscb_pratik}, it was assumed that before crossing, pedestrians from one group form a complete graph -- being connected to each other by virtual connections or `edges'. During crossing, pedestrians from one group `cut' the edges between pedestrians from the other group. After crossing, the edges which are never cut, are left as the stripes. So essentially, pedestrians from the same stripe from a particular group follow the same crossing pattern with respect to agents from the other group - this is the key idea that we explored while constructing the matrix $M$ or $\text{sgn}(M)$.

\begin{figure}[h!]
    \centering
    \includegraphics[width=\textwidth]{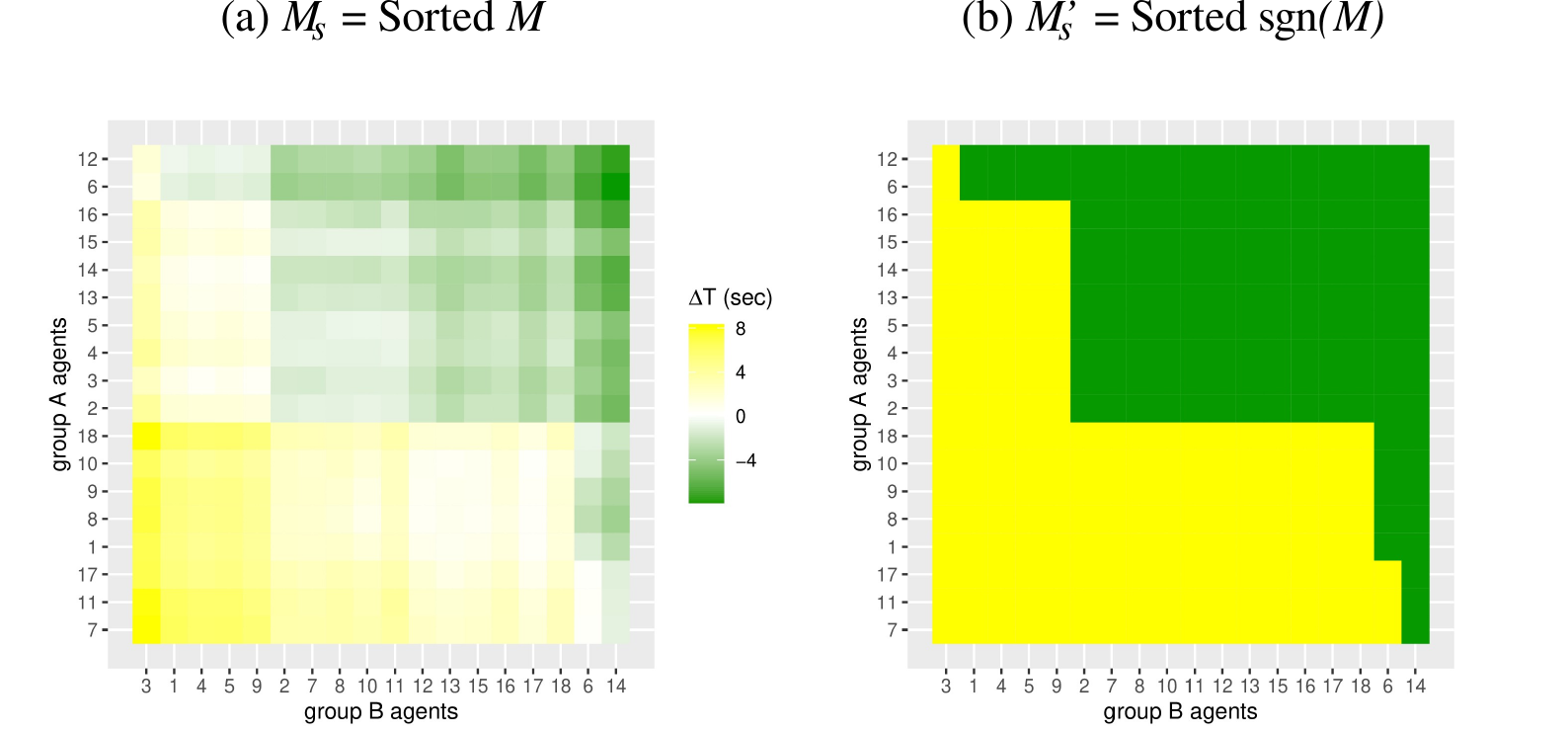}
    \caption{Sorted matrices (a) $M_s$ and (b) $M'_s$ for the configuration of crossing flows that has been shown in Figure \ref{matrices}. The matrices $M_s$ and $M'_s$ are obtained by sorting the rows and columns of the matrices $M$ and $\text{sgn}(M)$ respectively. Presence of stripes could easily be detected by looking at the 'knee' points of the matrices.}
    \label{matrices_sorted}
\end{figure}

To obtain the number of stripes that is formed from a group, we first count the number of unique patterns present in the matrix $M$ or $\text{sgn}(M)$. For this purpose, it is useful if we sort the rows and columns of these two matrices in either ascending or descending order of the sum of values of the rows or columns. We define, $M_s$ as the sorted matrix of $M$ and $M_s'$ as the sorted matrix of $\text{sgn}(M)$. The ordering strategy (ascending or descending) does not effectively change the crossing behavior of the agents, as represented by the matrices. For the same reason, sorting first along rows or columns does not matter. In Figure \ref{matrices_sorted} we have shown the matrices $M_s$ and $M_s'$, which are the sorted matrices of $M$ or $\text{sgn}(M)$ respectively as shown in Figure \ref{matrices}.

The sorting procedure results in segregation of yellow and green regions in the matrices and groups the agents with unique patterns together. By looking at the boundary of these two regions and the unique patterns we can detect the presence of stripes. From Figure \ref{matrices_sorted}, we can see that agents 14 and 3 from group $B$ crosses first and last, respectively, with respect to all the agents in group $A$. So both of them creates a stripe of size (technically not a stripe). Next we see that agents 7, 11 and 17 from group $A$ has a unique pattern and crosses first with respect to all agents in group $B$, except agent 14. So we can say that agents 7, 11 and 17 from group $A$ form a stripe of size 3 and cross after agent 14 of group $B$. This stripe is followed by agent 6 from group $B$ and a stripe from group $A$ of size 5 consisting of agents 1, 8, 9, 10 and 18. We can proceed our pictorial analysis in this way and obtain the stripe formation information, which are summarised in Table \ref{table_stripes}.\begin{table}[h!]
    \centering
    \begin{tabular}{|c|c|c|c|}
         \hline 
         Order of&Group&Size of&Agents in\\
         crossing&&the stripe&the stripe  \\
         \hline
         9&B&1&3\\
         \hline 
         8&A&2&6, 12 \\
         \hline 
         7&B&4&9, 5, 4, 1\\
         \hline 
         6&A&8&2, 3, 4, 5, 13, 14, 15, 16\\
         \hline 
         5&B&11&18, 17, 16, 15, 13, 12, 11, 10, 8, 7, 2 \\
         \hline 
         4&A&5&1, 8, 9, 10, 18 \\
         \hline 
         3&B&1&6 \\
         \hline 
         2&A&3&7, 11, 17\\
         \hline 
         1&B&1&14 \\
         \hline 
    \end{tabular}
    \caption{Information about the formation of stripes that could be obtained from the matrices shown in Figure \ref{matrices_sorted}}
    \label{table_stripes}
\end{table} So we can see that the matrices constructed by us not only provide the structural information about the stripes, but also the temporal ordering in which the stripes do cross each other and eventually pass the crossing region. 

\begin{figure}[h!]
    \centering
    \includegraphics[width=0.85\textwidth]{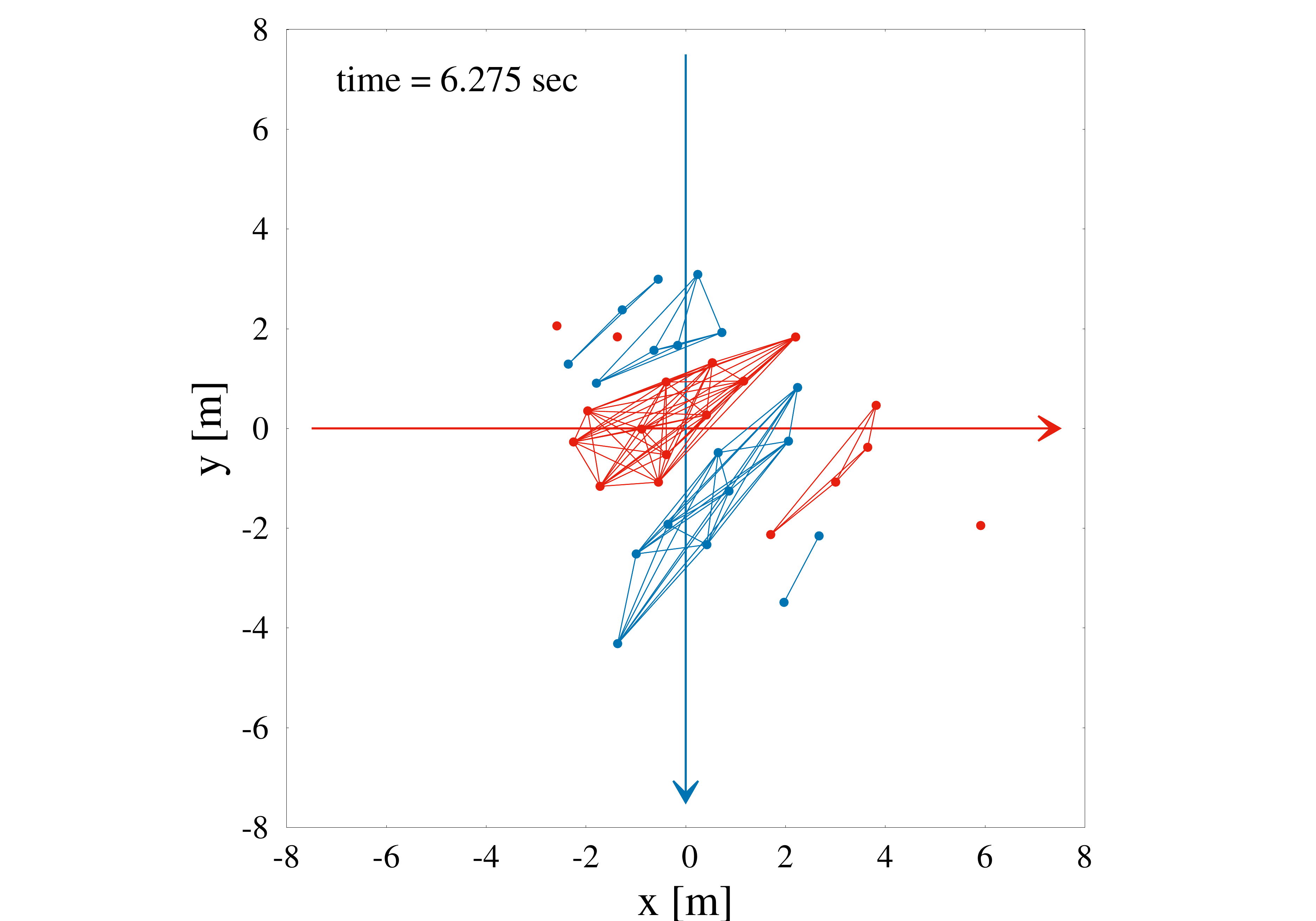}
    \caption{Geometric construction of the stripes as obtained from the edge-cutting algorithm applied on the trial of crossing flows for which the matrices are shown in Figures \ref{matrices} and \ref{matrices_sorted}. The blue and red dots denote the agents from group 1 and group 2,respectively, whose directions of motion are represented by arrows of color blue and red, respectively. The theoretical crossing angle of this trial is 90\degree. The stripes that are formed could be seen as complete graphs of agents connected to each other by virtual bonds, called edges. The instant of this snapshot is 6.275 sec after the agents started to walk. The entire video of this process is available in the Supplementary Materials. This figure is to be compared with Table \ref{table_stripes}.}
    \label{edgecutting}
\end{figure}

Our proposed method of using crossing matrices to obtain detailed information about the formation of stripes could be verified with the already established edge-cutting algorithm \cite{ploscb_pratik}. In Figure \ref{edgecutting}, we show the geometric construction of the stripes as obtained from the edge-cutting technique, when applied on the trial of crossing flows for which the matrices are shown in Figures \ref{matrices} and \ref{matrices_sorted}. Comparing this figure with the information that we get from the matrices as summarised in Table \ref{table_stripes}, we conclude that our matrix method provides the same information in much more elegant fashion and without any complicated calculations, as compared to the edge-cutting algorithm.

\subsection{Analytical predictions from the geometric model}\label{sec:res_gemoteric}

Here we discuss the analytical predictions of the number of stripes $\hat{n}$ and the crossing time $\hat{\tau}^*$ that could be made from our elliptical model (Section \ref{sec:geometric}), and compare them with the exact values obtained from the edge-cutting algorithm (EC) \cite{ploscb_pratik}. The main idea behind our model is to consider a situation before the two ellipse-shaped groups start to cross, and then find crossing angle dependent expressions for $\hat{n}$ (Eq. \ref{eq:stripe_number}) and $\hat{\tau}^*$ (Eq. \ref{eq:crossing-time}). So, these estimations are like predictions for the two quantities even before the crossing has taken place. In this analysis, we consider the time frame $\tau-1$ second to estimate the values of $a$, $b$ and $V$ from the experimental data. According to our model $a$ and $b$ are the semi-axes of the ellipses, but in reality the groups are indeed not elliptical. However, we can still consider $a$ and $b$ as the half-sizes of the group along $\overrightarrow{V}$ and along perpendicular to $\overrightarrow{V}$.



In our simplified elliptic model of crossing flows we assume that $a$, $b$ and $V$ do not depend on the crossing angle. Because of this assumption we calculate these values from all the trials, and consider their global averages to find estimates of $\hat{n}$ and $\hat{\tau}^*$. We found $a=4.957 \pm 0.667 m$, $b=3.686 \pm 0.779 m$  $V=1.113 \pm 0.237 m/s$ and $\overline{d}_{min} = 0.853 \pm 0.091 m$, hence $S = 1.172 m^{-1}$

\begin{figure}[h!]
	\centering
	\includegraphics[width=1.0\textwidth]{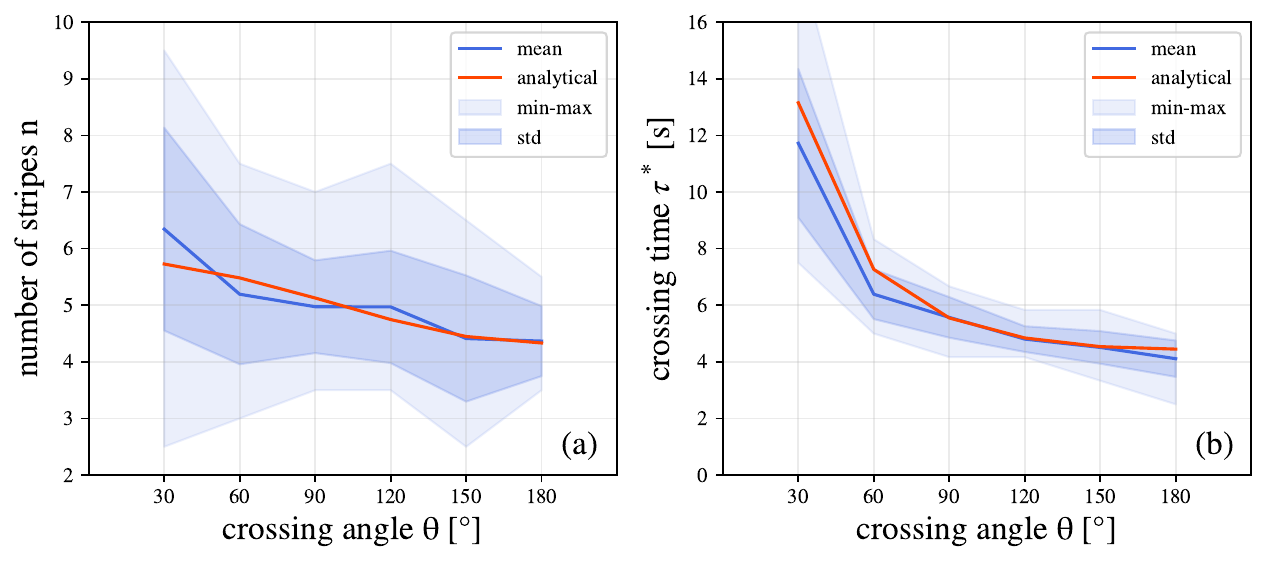}
	\caption{Variations of (a) number of stripes $n$ and (b) crossing time $\tau^*$ as a function of crossing angle $\theta$. In both the cases, the blue lines indicate the mean values of these quantities as obtained from the edge-cutting algorithm. The red lines indicate our analytically obtained results from Eqs. (\ref{eq:stripe_number}) and (\ref{eq:crossing-time}) for $n$ and $\tau^*$ respectively.}
    \label{fig:interaction-time-stripes} 
\end{figure}

The number of stripes $\hat{n}$ estimated using Eq. (\ref{eq:stripe_number}) are presented in Figure \ref{fig:interaction-time-stripes}(a). Despite some discrepancies, the variation of $\hat{n}$ with the crossing angle $\theta$ remains largely within the standard deviations of the experimental results obtained via the edge-cutting algorithm. This suggests that, although our elliptical model is deliberately minimalistic, it serves as a reasonable approximation. However, capturing finer details of the phenomenon may require more sophisticated modeling approaches.


In contrast, the estimated crossing time $\hat{\tau}^*$, calculated using Eq. (\ref{eq:crossing-time}), demonstrates notably higher accuracy with the experimental results, as shown in Figure \ref{fig:interaction-time-stripes}(b). The systematic variation of $\hat{\tau}^*$ with the crossing angle $\theta$ closely aligns with the experimental observations. This level of agreement is particularly impressive given the simplicity of the model in capturing the dynamics of a complex scenario.

\subsection{Group elongation}\label{sec:elongation}

\begin{figure}[h!]
\centering
\includegraphics[width=\textwidth]{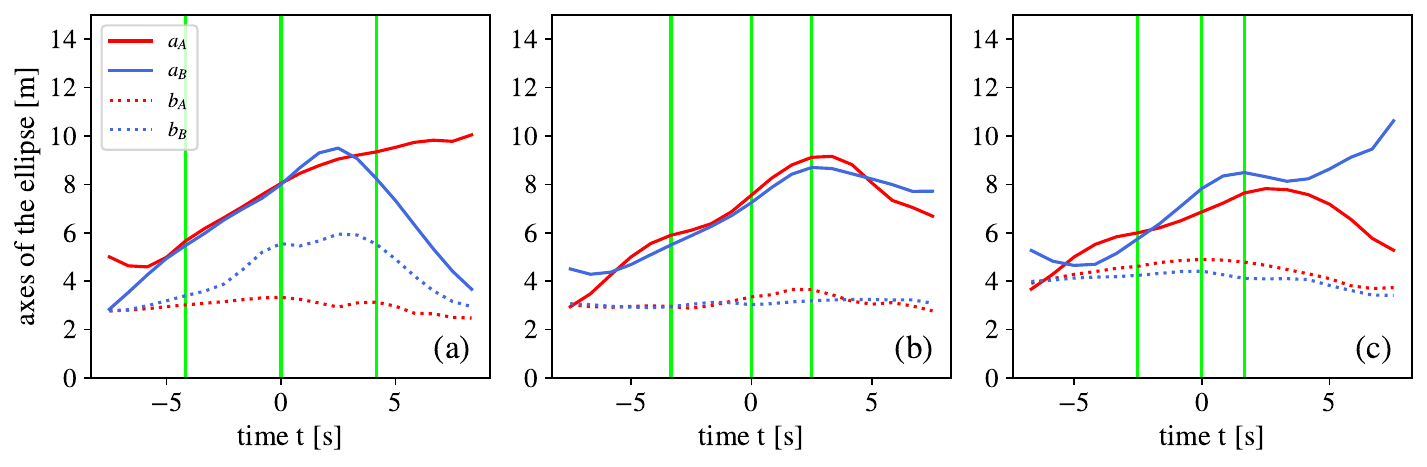}
\caption{Time evolution of axes $a$ and $b$ of of the ellipses for both the groups, shown for 3 typical trials with crossing angles (a) $\theta=60 \degree$, (b) $\theta = 90 \degree$ and (c) $\theta=120 \degree$ respectively. The solid and dotted lines represent $a$ and $b$ respectively. Green vertical lines on these panels denote: $\tau_1$, $\tau_2$ and $\tau_3$.}	
\label{fig:crossing-ab(t)}
\end{figure}

In our elliptical modeling approach for crossing groups, we assumed that the shape of the groups undergoes deformation during the crossing process. To gain deeper insight into this dynamic behavior, we compute $a(t)$ and $b(t)$ separately for both groups as functions of time $t$. These time-dependent variations are presented in Figure \ref{fig:crossing-ab(t)}(a)--(c) for three representative trials from the crossing flows dataset. We observe that $b(t)$ remains nearly constant, while $a(t)$ gradually increases until $\tau_3$, followed by a decrease. This indicates that the groups primarily elongate along the direction of motion, with little to no expansion in the perpendicular direction.

Interestingly, Figures \ref{fig:crossing-3-H}(d)--(i) create the visual impression that the groups are expanding perpendicularly to their movement. In reality, however, the elongation occurs along the movement direction. This visual illusion arises due to the use of a moving reference frame.

\begin{figure}[h!]
\centering
\includegraphics[width=1.0\textwidth]{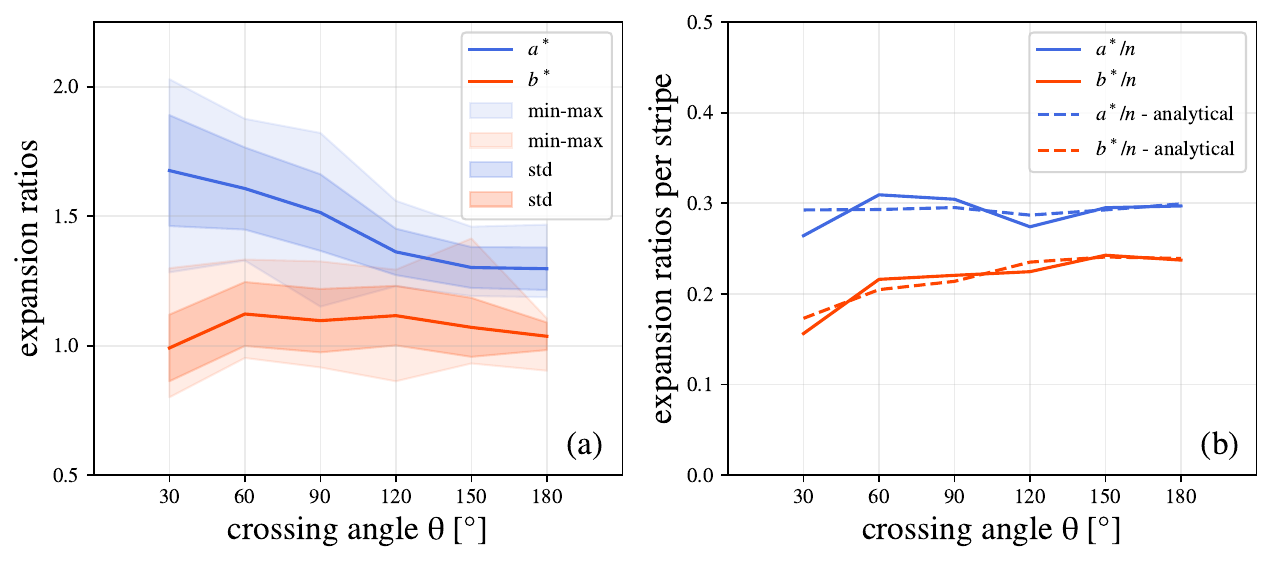}
\caption{Variation of (a) expansion ratios $a^*$ and $b^*$, and (b) average expansion ratios $a^*/n$ and $b^*/n$ as a function of crossing angle $\theta$. In (b), the solid and dashed lines indicate the values obtained from the edge-cutting algorithm and our geometric model (using Eq. (\ref{eq:stripe_number})) respectively.} 
   \label{fig:CM-expansion} 
\end{figure}

To quantify this deformation, we calculate the ratios $a^*=a(\tau_3)/a(\tau_1)$ and $b^*=b(\tau_3)/b(\tau_1)$, which represent the expansions in the parallel and perpendicular directions, respectively. The variations of $a^*$ and $b^*$ with respect to the crossing angle are shown in Figure \ref{fig:CM-expansion}(a). It is evident that $b^*$  remains close to 1 across angles, while $a^*$ tends to be significantly larger.

Upon comparing Figure \ref{fig:CM-expansion}(a) with Figure \ref{fig:interaction-time-stripes}(a) we can see that $a^*(\theta)$ has a strong correlation with $\hat{n}(\theta)$. Motivated by this, we proceed to estimate the average expansion of the groups per stripe as $a^*/\hat{n}$ and $b^*/\hat{n}$, and shown them in Figure \ref{fig:CM-expansion}(b). It is evident that the average elongation per stripe in the parallel direction of motion remains approximately constant across angles, whereas the elongation in the perpendicular direction increases with angle. This observation suggests an empirical relationship: \begin{align}
    &a^*\approx 0.3\hat{n}\\
    &a^*\approx 0.6\sqrt{a^2 \sin^2\frac{\theta}{2} + b^2 \cos^2\frac{\theta}{2} }
\label{eq:elongation_empir}
\end{align}This is a particularly interesting result: it indicates that the elongation ratio $a^*$ is, on average, proportional to the effective height $h$ of the group, which in turn is directly linked to the average number of stripes $\hat{n}$.

This result highlights a fundamental connection between the internal deformation of pedestrian groups and the large-scale structure of emergent flow patterns. It suggests that the extent to which groups stretch along their direction of motion (quantified by $a^*$) directly governs the number of stripes $\hat{n}$ that form in the crossing region. This understanding not only supports the validity of our elliptical model but also offers a simple yet powerful way to predict stripe formation based on group geometry. It points toward a broader principle in crowd dynamics: that collective structure can emerge as a natural consequence of individual group-level deformation.

\subsection{Observing stripes from barycentric approaches}\label{sec:observe}

Using our barycentric approaches, we successfully observe the well-known phenomenon of stripe formation, which consistently occurs perpendicular to the bisector of the crossing angle across all scenarios. Previous studies have relied on computationally intensive methods such as those in \cite{ploscb_pratik,worku2024detecting} to detect such stripes in the crossing region. In contrast, our approach offers an alternative perspective by employing barycentric reference frames, which enables the detection of stripes through visual inspection of transformed trajectories, without the need for complex calculations.

When pedestrian trajectories are linearly transformed into the global barycenter reference frame, the direction of movement of both groups becomes aligned along the horizontal axis. Consequently the stripes, which were originally perpendicular to the bisector in the static frame $F$, now appear aligned with the direction of motion in the transformed frame. Figure \ref{fig:crossing-3-H} illustrate this transformation, where the reference frame is rotated so that the velocity of the global barycenter points vertically upward. This rotation causes the relative motion of the two groups to unfold horizontally, making the stripes appear horizontal in the visualizations. This effect is clearly illustrated in Figure \ref{fig:crossing-3-H}(d)–(l) where the alignment of stripes becomes immediately visible.


One of the key advantages of this approach is its ability to capture multiple dynamical features—such as stripe formation and group deformation within a single visualization. While the raw trajectories shown in Figures \ref{fig:crossing-3-H}(a)–(c) offer limited understanding of the internal structure of the flow, their transformed counterparts reveal emergent spatial organization and group-level interactions with much greater clarity. Although the precision of our method does not match the strict quantitative accuracy of the edge-cutting algorithm (EC) from \cite{ploscb_pratik}, it compensates by offering a compact, interpretable, and visually effective representation of collective movement dynamics.

\subsection{Results from BCP and BCS methods}\label{sec:res_bc}

Here we summarize the results for obtained number of stripes using BCP and BCS methods described in section \ref{sec:number_of stripes}. In the case of the BCP method, we select the optimal combination of parameters—threshold type, time points, and aggregation function—for each crossing angle individually to minimize the prediction error. In contrast, the BCS method uses a fixed time point $\tau_{2.5}$ and an absolute threshold $T_a$ across all angles, offering a simpler yet reasonably accurate alternative. The only thing that differed between different crossing angles was the threshold $T_a$. We also report the Pearson's correlation coefficient $R$ between our obtained results and the results obtained from Edge-cutting algorithm \cite{ploscb_pratik}.

\begin{figure}[h!]	
\centering
\includegraphics[width=1.0\textwidth]{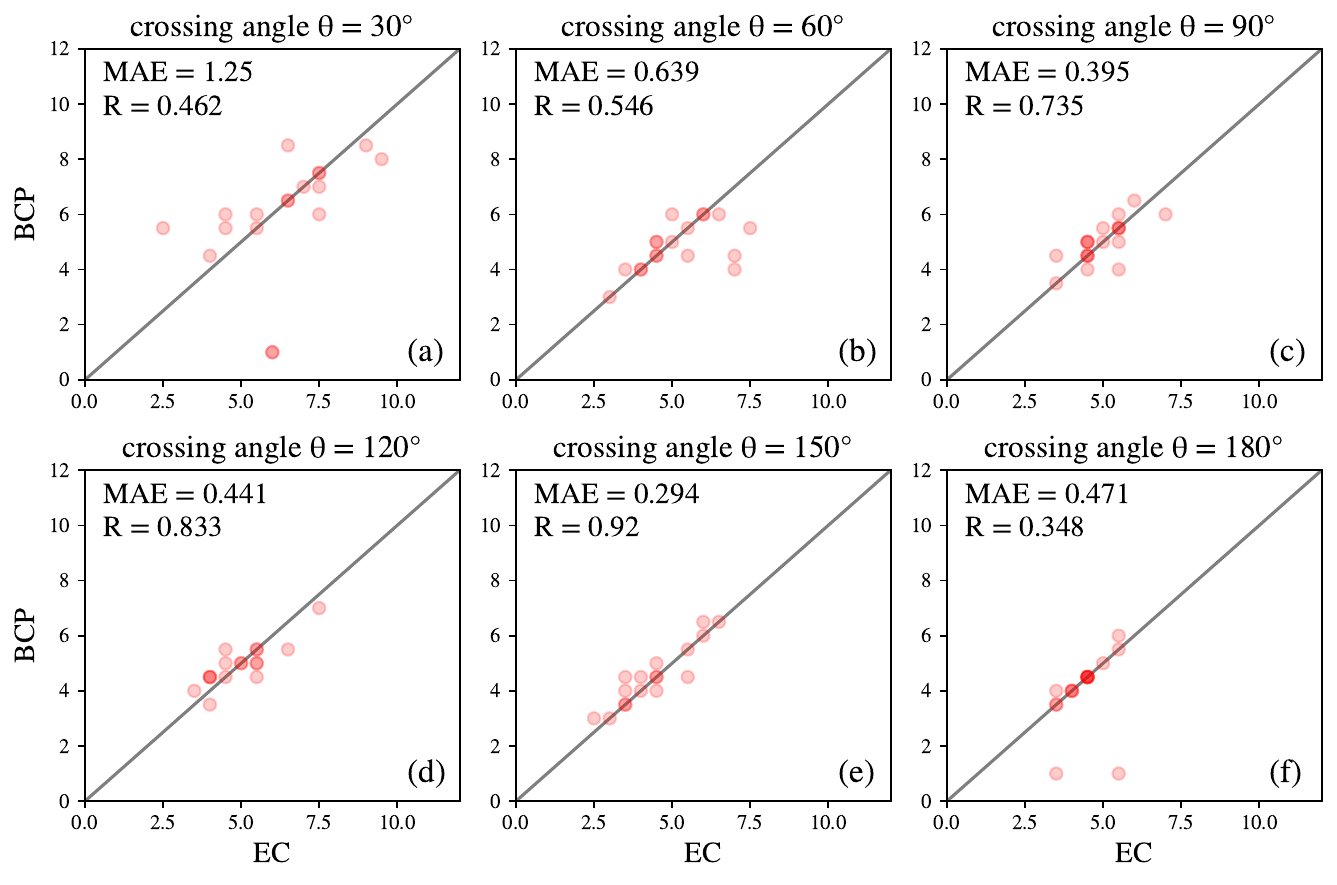}
\caption{Comparison of results obtained for number of stripes between edge-cutting (EC) algorithm and precise barycentric (BCP) method for different values of crossing angle $\theta$. For each case, the values of mean absolute error (MAE) and pearson's correlation coefficient $R$ are also reported.}     
\label{fig:BCP}
\end{figure}

\begin{figure}[h!]
\centering
\includegraphics[width=1.0\textwidth]{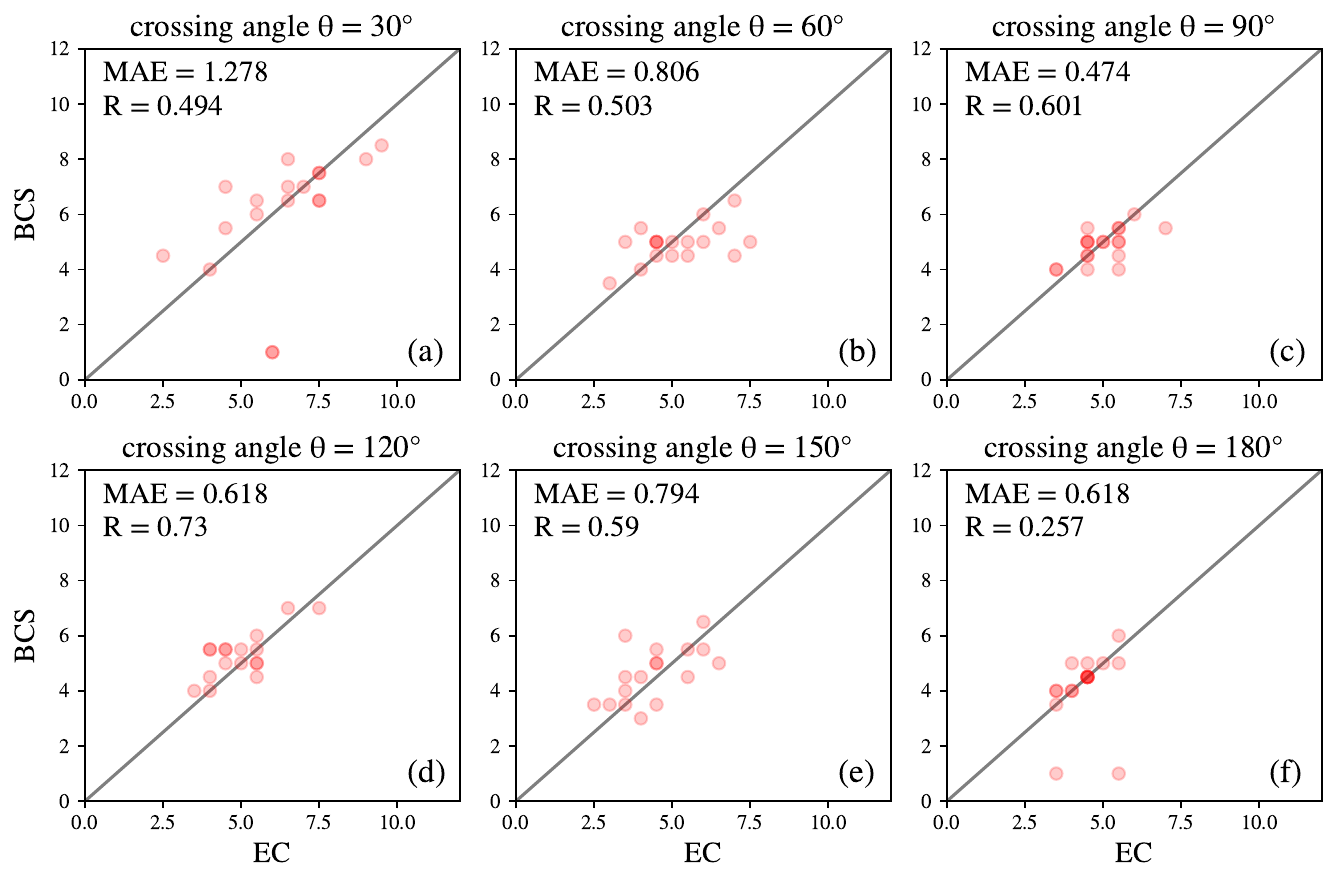}
\caption{Comparison of results obtained for number of stripes between edge-cutting (EC) algorithm and simplified barycentric (BCS) method for different values of crossing angle $\theta$. For each case, the values of mean absolute error (MAE) and pearson's correlation coefficient $R$ are also reported.}    
\label{fig:BCS}
\end{figure}

The results from the BCP method are presented in Table \ref{tab:BCP}, with a comparison to experimental data shown in Figure \ref{fig:BCP}. The method performs reasonably well for crossing angles between $90\degree$ and $180\degree$, while its accuracy is lower for $30\degree$ and $60\degree$, though it still offers some utility in these cases. It is important to note that the relatively low correlation at $180\degree$ is primarily due to the presence of two outliers. On the other hand, the estimations from BCS method is noticeably less precise, as summarized in Table \ref{tab:BCS}, and visually demonstrated in Figure \ref{fig:BCS}.

\begin{table}[h!]
    \centering
    \begin{tabular}{|c|c|c|c|c|c|}
         \hline 
         Crossing angle $(\theta$) & Type & Subset & Function & $T_z$ & MAE\\
         \hline
         30\degree & absolute & $\tau_{1.5}$, $\tau_{2.5}$ $\tau_{3}$ & min & 0.43m & 1.25\\
         \hline
         60\degree & relative & $\tau_{2}$, $\tau_{2.5}$ & floor & 0.318 & 0.639\\
         \hline
         90\degree & absolute & $\tau_{2}$, $\tau_{3}$ & min & 0.527m & 0.395\\
         \hline
         120\degree & absolute & $\tau_{2}$, $\tau_{2.5}$, $\tau_{3}$ & floor & 0.512m & 0.441\\
         \hline
         150\degree & relative & $\tau_{1}$, $\tau_{2}$ , $\tau_{3}$, & min & 0.263 & 0.294\\
         \hline
         180\degree & relative & $\tau_{1.5}$, $\tau_{2}$, $\tau_{2_5}$ & ceil & 0.433 & 0.471\\
         \hline 
    \end{tabular}
    \caption{Table shows results obtained from the precise barycentric method (BCP).}
    \label{tab:BCP}
\end{table}

\begin{table}[h!]
    \centering
    \begin{tabular}{|c|c|c|c|c|c|}
         \hline 
         Crossing angle ($\theta$) & $T_a$ & MAE\\
         \hline
         30\degree  & 0.592 & 1.278\\
         \hline
         60\degree  & 0.605 & 0.806\\
         \hline
         90\degree  & 0.642 & 0.474\\
         \hline
         120\degree  & 0.477 & 0.618\\
         \hline
         150\degree  & 0.369 & 0.794\\
         \hline
         180\degree  & 0.349 & 0.618\\
         \hline 
    \end{tabular}
    \caption{Table shows results obtained from simplified barycentric method (BCS).}
    \label{tab:BCS}
\end{table} 

While the average number of stripes estimated by both methods remains close, notable discrepancies persist when compared to edge-cutting algorithm. It is important to acknowledge that the edge-cutting algorithm serves as a ground truth, offering exact stripe counts, whereas the barycentric methods presented here is an approximation. Although it lacks the precision of the original method, it achieves reasonably accurate results with significantly lower computational effort. This trade-off makes it a promising tool for analyzing large datasets or for calibrating mechanistic models, even if it may not be ideal for evaluating individual trials with high accuracy. In summary, the barycentric perspective offers a compelling way to detect the presence and orientation of stripe patterns, clearly showing their alignment perpendicular to the crossing angle, but falls short in providing precise quantification, a task better suited to more exact methods. It is however possible, that the direction indicated by these methods has potential for further exploration. There are many other potential possibilities of exploring trajectories in the barycentric reference frames.

\section{Conclusions}\label{sec:conclusion}

In this study, we introduced two methods of experimental data analysis and one model based on geometric properties of the crossing pedestrian flows. The first data analysis method is a matrix-based method, which extracts the temporal ordering of individual crossings by analyzing the time differences at points of minimal distance between trajectory pairs from opposing groups. This method constructs a matrix that encodes the relative crossing sequences and enables exact detection of stripe structures. While computationally less demanding than the edge-cutting algorithm, it provides comparable results, making it a practical tool for large-scale data analysis. The second method of data analysis is based on barycentric reference frames. It presents interesting perspective, and has potential for future development, but unfortunately lacks accuracy of a matrix-based method. 

The geometric model based on elliptical approximations of the crowd shape, helps to understand and quantify some crucial aspects of the dynamics. This model allows for analytical estimation of key quantities viz. the number of stripes and the total interaction time, as functions of the crossing angle. Despite its simplicity, the model aligns well with experimental observations and captures the underlying geometry of group deformation during crossing. Together, these two methods and the model offer efficient and interpretable strategies for linking microscopic pedestrian interactions to emergent macroscopic flow patterns.


An important empirical observation in our study is the expansion of pedestrian groups during crossing, with the degree of expansion varying with the crossing angle. Notably, the parallel expansion ratio $a^*$ was found to be approximately proportional to the vertical cross-section of the group in the global reference frame, suggesting a strong link between group geometry and emergent stripe structure. Furthermore, we observed that both the number of stripes and the total interaction time decrease systematically with the crossing angle, consistent with that reported in \cite{ploscb_pratik}. As shown in Figure \ref{fig:interaction-time-stripes}, model explains averages of observed quantities with o very good level of agreement. However, purely geometric parameters do not fully explain the observed expansion and interaction time, indicating that additional dynamic factors may be involved.



Beyond methodological contributions, our work offers broader implications for urban planning, crowd management, and simulation-based modeling. Stripe formation and group-level dynamics are crucial for understanding and optimizing pedestrian flows in high-density urban environments, such as train stations, intersections, and evacuation scenarios. The matrix method, in particular, enables real-time detection of collective structures with minimal computational overhead, supporting large-scale crowd monitoring and control. On the other hand, the elliptical model provides quantitative benchmarks for calibrating agent-based or continuum simulations, ensuring that such models reproduce realistic timing and spatial structure. These methods can support the design of infrastructure and policies that promote smoother pedestrian interactions and reduced congestion.


\textit{Future research}: Building on these findings, we propose two hypotheses as avenues for further exploration. First, if continuous streams of pedestrians were to cross instead of finite groups, we expect the emergence of traveling stripe waves in the interaction region, with movement velocity approximated by the vector average of the pedestrian velocities. This is just a generalization of observations made for finite groups. In case of finite groups we also observed striped pattern to travel with the velocity being an average of the vectors of all pedestrians. Namely, velocity of the global barycenter. However, finite groups created only small pattern of stripes, and continuous stream will produce continuous traveling stripe pattern. 
Second, as the thickness of the crossing beams increases, so should the expansion region, possibly leading to a transition from structured stripes to a disordered or turbulent regime beyond a critical threshold. This is because in small region of finite groups we have a lot of space to move and eventual perturbations do not add up much. In larger crossing region we would have less freedom for the agents to move, and perturbations of movement could build up and accumulate in the fashion similar to traffic jam models. 
These conjectures highlight the rich, complex dynamics that may arise in more persistent or larger-scale flows and motivate the need for experiments and studies under controlled conditions, followed by modeling attempts. Future work may also aim to incorporate adaptive behavioral rules and explore boundary effects, further enhancing the predictive power and applicability of these models in diverse real-world contexts.

\bibliographystyle{elsarticle-num} 
\bibliography{references_A_3}

	
\end{document}